\documentclass[conference]{IEEEtran}
\IEEEoverridecommandlockouts
\usepackage{cite}
\usepackage{amsmath,amssymb,amsfonts}
\usepackage{algorithmic}
\usepackage{graphicx}
\usepackage{float}  %设置图片浮动位置的宏包
\usepackage{subfigure}  %插入多图时用子图显示的宏包
\usepackage{textcomp}
\usepackage{enumerate}
\usepackage{xcolor}
\def\BibTeX{{\rm B\kern-.05em{\sc i\kern-.025em b}\kern-.08emT\kern-.1667em\lower.7ex\hbox{E}\kern-.125emX}}
\begin{document}

\title{An Overview of Resource Allocation in Integrated Sensing and Communication\\

}

\author{\IEEEauthorblockN{ Jinming Du, Yanqun Tang{*}, Xizhang Wei, Jiaojiao Xiong, Jiajun Zhu, Haoran Yin, Chi Zhang, Haibo Chen}
\IEEEauthorblockA{\textit{School of Electronics and Communication Engineering, Sun Yat-sen University, Shenzhen, China} \\
Corresponding Author: tangyq8@mail.sysu.edu.cn
%email: {dujm3@mail2.sysu, tangyq8@mail.sysu, weixzh7@mail.sysu, xiongjj7@mail2.sysu, zhujj59@mail2.sysu, yinhr6@mail2.sysu}.edu.cn
}
}
\maketitle

\begin{abstract}
Integrated sensing and communication (ISAC) is considered as a promising solution for improving spectrum efficiency and relieving wireless spectrum congestion. 
This paper systematically introduces the evolutionary path of ISAC technologies, then sorts out and summarizes the current research status of ISAC resource allocation.
From the perspective of different integrated levels of ISAC, we introduce and elaborate the research progress of resource allocation in different stages, namely, resource separated, orthogonal, converged, and collaborative stages.
In addition, we give in-depth consideration to propose a new resource allocation framework from a multi-granularity perspective.
Finally, we demonstrate the feasibility of our proposed framework with a case of full-duplex ISAC system.

\end{abstract}


\section{Introduction}
Over the past few decades, wireless communications and radar sensing have each evolved by leaps and bounds. Driven by the six-generation (6G) complex application scenarios, sensing and communication (S\&C) will be highly coupled in the future. Integrated sensing and communication (ISAC) has been proposed as a critical technology for 6G wireless networks and radar systems, which enables simultaneous sensing and communication transmission operations through spectrum coexistence and hardware sharing. The trend of 6G to higher frequency bands and the similarities between S\&C in hardware architectures, channel characteristics, and signal processing have contributed to the development of ISAC techniques.

With the ever-increasing demand for bandwidth and data rates, ISAC has garnered significant attention from academic and industry communities\cite{9737357}. Early works on ISAC mainly focused on the coexistence of radar and communication systems when the spectrum overlaps, emphasizing the development of effective interference management technologies. To improve hardware efficiency, the literature has been trying to consider the integration of both functions into a single system. Inspired by the widespread use of multiple access technologies, ISAC allows the allocation of non-overlapping wireless resources for the two functions through a single transmitter. As a step further, waveform design is gradually becoming the mainstream of ISAC research, where a fully unified waveform allows more efficient use of wireless resources and provides the maximum integration gain.

From the perspective of the existing development stages, resource allocation in ISAC can be divided into four stages, namely resource separated, resource orthogonal/non-overlapping, resource convergence/overlapping, and resource collaborative networked stages. 
The scheduling of S\&C functions on orthogonal wireless resources can be implemented in time, frequency, space, and code domains. Based on the design priority, the resource converged stage can also be subdivided into the communication-centric design, radar-centric design, and joint design.
Although ISAC has been well investigated from various aspects in recent years, the model of resource allocation is still relatively simple and the application scenarios are limited, thus a unified resource allocation architecture is in urgent need of being widely explored.

This paper aims to categorize and describe the existing major ISAC resource allocation schemes based on the degree of integration, and further analyzes and compares their advantages and disadvantages.
In addition, we propose a new granularity-based resource allocation framework that aims to achieve optimal allocation from multiple layers, domains, dimensions, and scales.
Finally, we outline and enumerate the practical challenges in this area and identify future research directions.

\section{The Development of ISAC}

\subsection{ISAC: From Coexistence to Integration}
The development process of ISAC can be divided into four stages\cite{liu2023seventy}. Firstly, in the initial stage, spectrum resources are shared among various S\&C systems and the main challenge is how to reduce interference among them\cite{7470514}. In the second stage, the originally separated S\&C systems are integrated in the same physical platform, in which each block of wireless resources is dedicated to a specific function.
In the third stage, S\&C can be simultaneously supported by a single system, using the same transmit waveform and a unified signal processing framework. This signal-level integration enabled the full reuse of wireless resources, and the waveform design, and signal processing algorithms are the research hotspot. In the final stage, S\&C can share a common network infrastructure, i.e., forming 6G base stations and terminals with both communication and sensing functions to build an ISAC network. The technical proposals concentrate on multi-point sensing and collaborative networking, which will provide firm support for emerging 6G applications. Based on the above four stages of development, ISAC technology will eventually build the endogenous sensing capability of 6G.

\subsection{The Development of Resource Allocation in ISAC}
Due to the limited hardware and wireless resources, it is necessary to make reasonable trade-offs based on specific metrics of S\&C performance, and then design an optimal resource allocation scheme. In the research of ISAC, the researchers often use traditional metrics such as Shannon capacity and bit error ratio, while the sensing selects metrics based on application scenarios, which can be broadly classified into three categories: detection, estimation, and recognition.

Early works on resource management in ISAC, merely focusing on unilateral optimization of communication or sensing performance, such as maximizing the communication throughput $C \left(p_r, p_c\right)$ within the radar service constraint\cite{8834831}, 
\begin{equation}
\begin{aligned}
&\max _{p_r, p_c}  \operatorname\;C \left(p_r, p_c\right) \\
&\text { s.t. : } \left\{\begin{array}{l}
0 \leq p_r \leq 1,0 \leq p_c \leq 1 \\
\operatorname{SINR}_r \left(p_r, p_c\right) \geq \kappa_r
\end{array}\right.
\end{aligned}
\end{equation}
where $\kappa_r$ is a lower bound on the signal to interference plus noise ratio (SINR) for radar, $0\leq p_r\leq1$ and $0\leq p_c\leq1$ are the power allocation coefficients to be determined for communication and radar, respectively.
In order to ensure a harmonious coexistence between the two functions, the resource competition between communication and sensing needs to reach a reasonable balance. 
%The classification of sensing methods in ISAC can be divided into pilot-based and echo-based sensing. For the pilot-based sensing case, \eqref{eq} shows the trade-off between communication data rate and channel state estimation when the power of $\gamma P$ and $(1-\gamma) P$ is allocated for communication and sensing, respectively.
%\begin{equation}
%\begin{aligned}
%(R, D)= & \left(\frac{1}{2} \log _{2}\left(1+\frac{\gamma P}{\sigma_{n}^{2}}\right)\right. ,\\
%& \left.\sigma_{x}^{2} \frac{\left(\gamma P+\sigma_{n}^{2}\right)}{\left(\sigma_{x}+\sqrt{(1-\gamma) P}\right)^{2}+\gamma P+\sigma_{n}^{2}}\right)\label{eq}
%\end{aligned}
%\end{equation}
The studies in \cite{8437621} reveal the fundamental capacity-distortion tradeoff in ISAC and provide theoretical support for resource allocation schemes. It is theoretically demonstrated that ISAC driven systems have further gains over the systems where communication and sensing work separately and independently. In \cite{9424454}, the authors first presented the Pareto performance boundary for ISAC systems from the perspective of multi-objective optimization. In \cite{9933849}, the authors provided a robust resource allocation scheme for secure communication in ISAC systems.

However, the current ISAC resource allocation model is still relatively simple and thus can only be applied to a limited number of scenarios. Designing novel metrics and optimization criteria to describe the joint performance and achieve optimal resource allocation will be one of the biggest challenges in ISAC.

\section{Overview of Resource Allocation in ISAC}
\begin{figure}[htbp]
\centerline{\includegraphics[width=0.5\textwidth,height=0.25\textwidth]{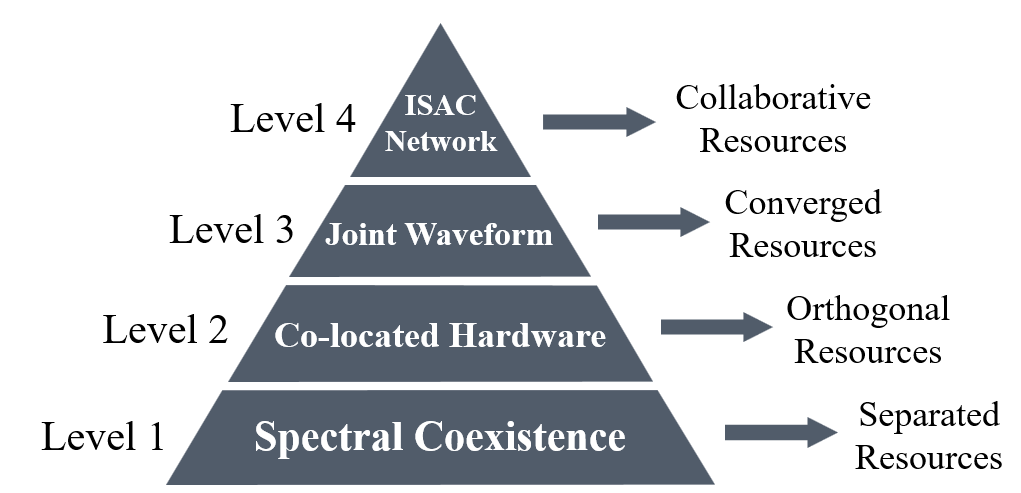}}
\vspace{-0.4em}
\caption{The evolution of ISAC technologies.}
\label{1}
\vspace{-1em}
\end{figure}

Following the development of the resource allocation methods, the current ISAC systems can be classified into the following four categories, as shown in Fig.~\ref{1}:
\begin{itemize}
\item Resource separated ISAC architecture;
\item Resource orthogonal ISAC architecture;
\item Resource converged ISAC architecture;
\item Resource collaborative ISAC network architecture.
\end{itemize}
In this section, we review and describe in detail the resource allocation characteristics for different levels of integration and then further analyze their advantages and disadvantages.
\subsection{Resource Separated ISAC Architecture}
Level 1 in Fig.~\ref{1} involves integrating a sensing system and a communication system only at the application layer to obtain an overall dual-function effect. 
In the point of spectral coexistence, both radar and communication systems have their own active transmitters, using completely different hardware architectures, and the most critical issue is the reduction or elimination of interference\cite{6558027}. 
%At first there was no cooperation between the two systems, focusing only on radar performance or only on communication performance.
For example, communication-centric approaches have been proposed that can counteract radar interference at the communication receiver \cite{8233171} or, taken directly at the transmitter to counteract radar-sensing interference with some prior information \cite{8332962}.
The two active systems can also cooperate in negotiating their respective transmission strategies and adjusting their detection strategies accordingly, which presupposes the existence of a fusion centre accessible to both systems\cite{7470514}. The presence of such accessible joint data centers eliminates interference more effectively, but increases hardware overhead.

\subsection{Resource Orthogonal ISAC Architecture}
It can be easily understood that S\&C can be scheduled on orthogonal/non-overlapping wireless resources so that it can mitigate interference between the two functions on co-located hardware. In this way, signals can be multiplexed in distinct dimensions such as time, frequency, space and code.
\paragraph{Time Division}
Time division multiplexing refers to performing the functions of radar detection and data communication respectively in different cycles without waveform redesign.
Previous researches can perform time division on frame, subframe, time slot and even symbol level\cite{7952964}, \cite{9528013}, \cite{2013Joint}, \cite{9728752}.
One of the most significant advantages of the time-division approach is that it is easy to implement and can be integrated into any system. However, the fact that only one function can be operated at a time makes it critical to allocate the time between the two functions fairly. 
The time allocation strategy can be adjusted according to dynamic demands. For example, in \cite{9528013}, with the increase of the time slot allocation ratio of radar function, the ranging performance of each vehicle is effectively improved. 

\paragraph{Frequency Division}
Frequency division multiplexing means that the radar and communication functions are implemented to transmit at different frequencies. The two functions are allocated to different subcarriers according to the channel conditions, different key performance indicators (KPIs) required for systems and power budget of transmitters \cite{9737357}.
For instance, Hossain et al. estimated the speed of a vehicle by using a vehicle-mounted radar with a carrier frequency of 77 GHz and used a carrier frequency of 5.9 GHz to transmit data to realize vehicle scheduling at intersections\cite{8251701}. The advantage of frequency division multiplexing is that the two functions are implemented simultaneously. However, the nonlinear distortion of the channel will change the actual frequency characteristics, which is easy to cause interference problems.

\paragraph{Spatial Division}
Spatial division multiplexing is a beam division system that divides the radar active phased antenna array into different subarrays, and S\&C are realized over orthogonal spatial resources\cite{9501235}. 
The researchers mainly consider sidelobe control schemes, interference performance analysis, and precoding waveform optimization. When the total energy is fixed and the active phased array antenna array is divided into different subarrays, the energy used for radar detection of the system is reduced, resulting in the degradation of radar performance. Therefore, it is very important to allocate spatial beam resources reasonably according to the needs of different scenarios.

\paragraph{Code Division}
Code division multiplexing has also emerged as a potentially applicable technology, where orthogonal sequences or quasi-orthogonal sequences are used to carry S\&C signals.
It can enable S\&C two functions to jointly use the time-frequency-space resources to improve spectrum resource utilization efficiency while sacrificing computational complexity.
Chen et al. proposed a new code division multiple access orthogonal frequency-division multiplexing (OFDM) waveform by multiplying the direct sequence spread spectrum code vector and information symbols, which uses the self-interference cancellation (SIC) scheme to alleviate mutual interference and achieve the mutual gain effect of S\&C performance\cite{9359665}.

\subsection{Resource Converged ISAC Architecture}
Resource orthogonal ISAC architecture is simple to implement and compatible with existing systems, but it does not fully utilize all resources in radar and communication. 
Resource converged ISAC architecture with joint waveform can be understood that radar and communication functions converge on a single waveform and the system transmits a shared waveform to achieve two functions\cite{9880774}, \cite{yin2023cyclic}.
The research focus of this architecture is waveform designs. According to the design priority and signal formats, resource converged ISAC architecture can be subdivided into communication-centric design, radar-centric design and joint design. 

\paragraph{Communication-Centric Design} 
Communication-centric design refers to modifying the original communication waveform to perform radar detection, in which case the communication functionality has priority. At present, the research can be mainly divided into strategies based on spread spectrum technology and OFDM. To obtain better communication performance, it is a good idea to use spread spectrum signals with good autocorrelation characteristics as the waveform of ISAC\cite{5776640}. OFDM is compatible with state-of-the-art standards, and the range and Doppler parameters can be readily obtained via fast Fourier transform, the use of OFDM for radar detection has recently received growing attention\cite{9729203}. Most OFDM researches focus on optimizing and improving the ambiguity function, autocorrelation function, peak-to-average envelope ratio and peak-to-average power ratio\cite{9266515}. Since the communication waveform is not customized for radar, its sensing performance is relatively less reliable, requiring complex signal processing techniques to compensate for performance losses.
\paragraph{Radar-Centric Design}  
Radar-centric design introduces differences in the radar waveforms and uses these differences to modulate communication information\cite{9127852}. 
Early design is to modulate communication information onto linear frequency modulation signal, and we can use the pulse starting frequency, step frequency, phase, pulse width, and other parameters to characterize communication information, which is time-frequency domain embedding\cite{8828030}. 
Furthermore, we can also embed communication information into the radar’s spatial domain, such as representing each communication information by the sidelobe level of the multiple input multiple output (MIMO) radar beam pattern or the antenna indices\cite{7347464}. 
As a step forward, the full-duplex (FD) technology has recently been introduced into the design. The radar pulse signal modulated by phase shift keying can be used to transmit low-rate commands and control information, and the waiting time of the pulse radar can be used to transmit high-rate communication signals\cite{9724187}.
The echo can even enhance sensing ability as long as self-interference is effectively suppressed.
Therefore, full duplex can significantly improve communication rates while reducing radar blind spots.

\paragraph{Joint Design} 
Although the communication-centric design and radar-centric design realize ISAC, they lack flexibility and fail to achieve scalable regulation of communication and perception functions. Therefore, joint design becomes a promising strategy, and the core problem is how to optimize the final transmitted signals to achieve a compromise between communication and sensing performance. 
Based on this, the design can be formulated as mathematical optimization problems. And the problems are often solved through convex optimization techniques, where the objective function is the sensing or communication performance metrics, with some constraints to ensure other functional performance\cite{9652071}. 
Some related works with different optimization goals correspondingly proposed different design algorithms, namely based on maximized mutual information design, waveform similarity design, estimation accuracy and analog array multi-beam optimization \cite{9540344},\cite{8386661}. 

%Taking analog array multi-beam optimization as an example, it is similar to the beamforming in traditional communication. For the original signal $s$, through the beamforming vector $\mathbf{w}_{T}$, the signal $\mathbf{x} = \mathbf{w}_{T}s$ is finally sent, and we need to design $\mathbf{w}_{T}$ so that it can have sufficient resources in the user direction and the sensing target direction\cite{8386661}.
%And the optimization problem can be written as:
%\begin{equation}
%\mathbf{w}_{T}=\sqrt{\rho} \mathbf{w}_{T, F}+\sqrt{1-\rho} e^{j \varphi} \mathbf{w}_{T, S}\mbox{,}
%\end{equation}
%where $\mathbf{w}_{T, F}$ and $\mathbf{w}_{T, S}$ corresponds to the fixed sub-beam and scanning sub-beam respectively.
%The symbol $\rho \in[0,1]$ is a scaling factor to control the weights assigned to S\&C functions, which may include some resources such as overall power budget constraint and per-antenna power budget constraint. Based on the actual needs of the system, we choose suitable $\rho$ and allocate power and other resources to each propagation path, thereby ensuring the desired S\&C performance.

\subsection{Resource Collaborative ISAC Network Architecture}
To realize the vision of 6G intelligent connectivity, it is necessary to explore the fundamental theory and resource allocation for the ISAC network collaboratively. 
In networked ISAC, information can be shared cross-layer, cross-module and cross-node to achieve fully integrated communication and sensing.
The two functions can mutually benefit each other, namely, communication-assisted sensing and sensing-assisted communication, which will bring significant performance improvement and substantial reduction in cost and energy consumption of the overall network system.

In communication-assisted sensing scenarios, aiming at optimizing the performance of mobile crowd sensing tasks, Li et al. proposed a joint optimal allocation scheme for S\&C resources in mobile edge networks\cite{9817050}. Long et al. used the synchronization functionality of communication to assist dual-base synthetic aperture radar imaging and eventually build a distributed radar network\cite{9961148}. In sensing-assisted communication scenarios, Jiao et al. realized the reconstruction of the environment and channel by radar detection and enhanced the communication rate by precise beamforming\cite{8680773}. Wang et al. envisioned a future working model of networked ISAC, where sensing nodes are endowed with new collaborative communication functions\cite{9668964}.

At present, the research on networked ISAC is still immature, and it will be a critical research direction in the future for effective network coordination, joint scheduling of resource allocation and interference management schemes to accomplish different sensing scenarios.

\section{A New Multi-granularity Resource Allocation Framework}
In the 6G era, there are a large number of nodes with co-existing S\&C functions, which poses a serious challenge for the traditional multi-dimensional resource allocation schemes. 
In this section, we attempt to consolidate the resource allocation in different stages of ISAC into a unified framework with two parts (inter-system and intra-system) and three layers (hardware layer, signal layer and application layer), and give an in-depth insight into future resource allocation from a unified multi-granularity-based perspective.
In order to further illustrate this framework, we provide an application case of resource allocation for FD-ISAC systems.
\subsection{The Multi-granularity ISAC resource allocation framework}
\begin{figure}[htbp]
\centerline{\includegraphics[width=0.45\textwidth,height=0.45\textwidth]{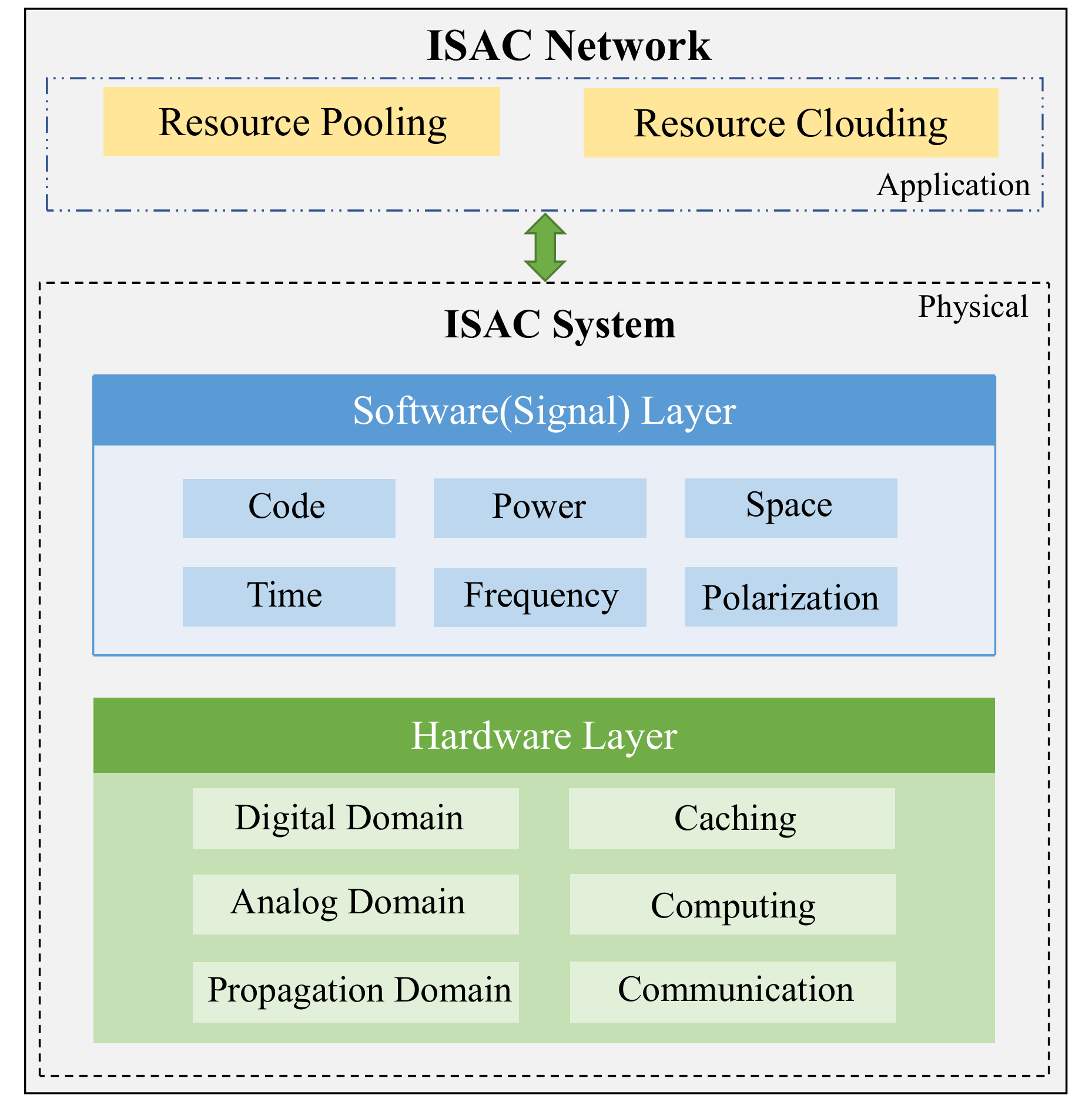}}
\vspace{-0.4em}
\caption{The multi-granularity ISAC resource allocation framework.}
\label{2}
\vspace{-1em}
\end{figure}
As shown in Fig.~\ref{2}, the multi-granularity resource allocation framework for ISAC can cover all the hardware resources and wireless resources in inter-system and intra-system (network) across the physical layer and application layer, where hardware and software layers are contained in the physical layer.

\begin{figure}[htbp]
\centerline{\includegraphics[width=0.36\textwidth,height=0.25\textwidth]{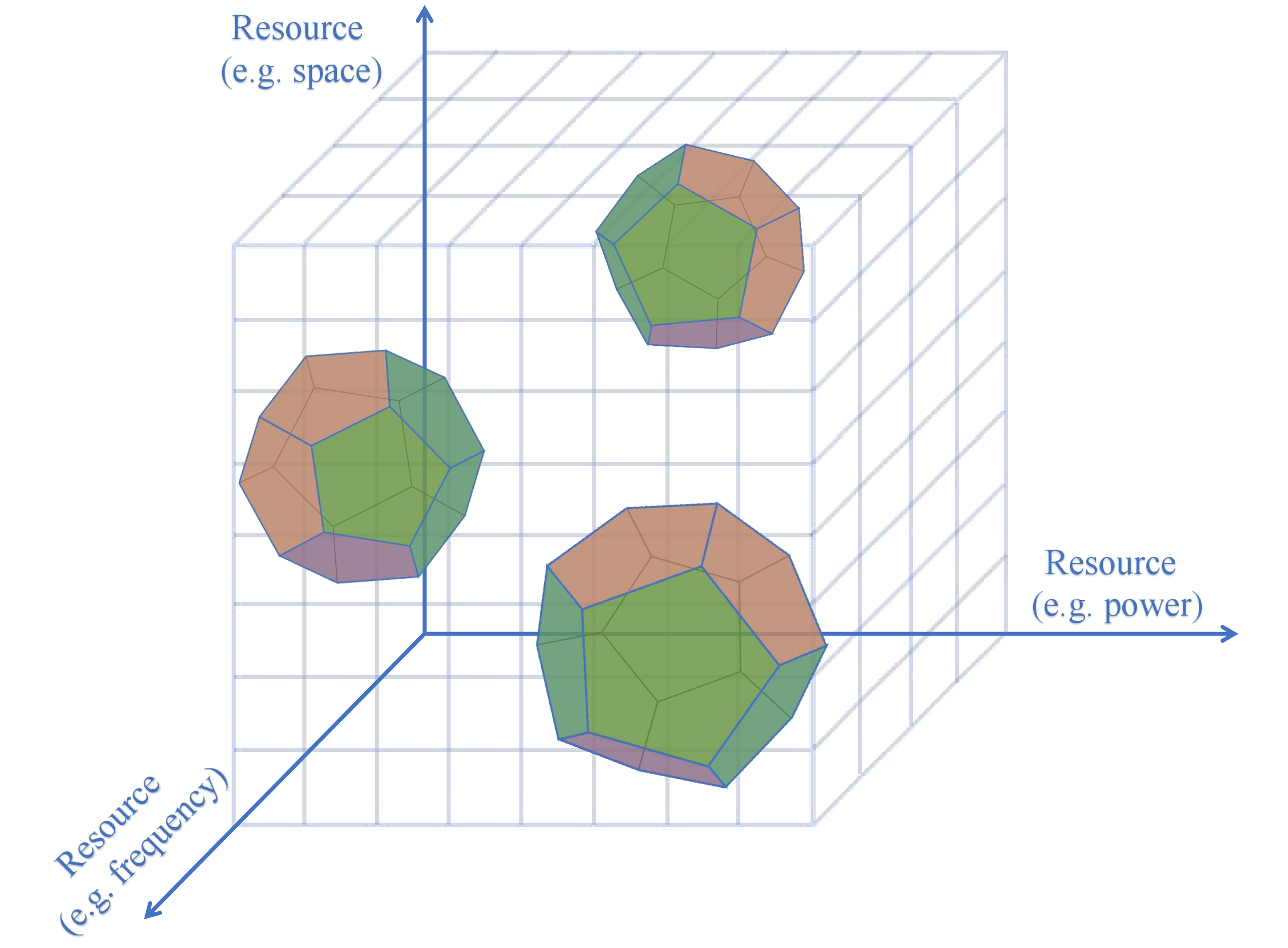}}
\vspace{-0.4em}
\caption{Multi-layer, multi-domain, multi-dimensional and multi-scale resource allocation forms based on granularity.}
\label{4}
\vspace{-1em}
\end{figure}

In the multi-granularity ISAC resource allocation framework, multi-granularity denotes a combination of multiple layers, domains and dimensions with different scales, as shown in Fig.~\ref{4}.
It is not difficult to imagine that in this multi-granularity model,  the optimal allocation of resource blocks matching different requirements will be scalable in multiple dimensions, or even be fluid form. 

Hardware resources include 3C resources of computing, caching and communication, which are used to perform sensing tasks and data processing with the cooperation of multiple domains such as propagation domain, analog domain, and digital domain. 
For the signal layer, it can be considered from different dimensions such as time, frequency, space, code, polarization and power, etc., to divide the resources as multiple scales like coarse-grained and fine-grained for flexible allocation.
Further, in the application layer, resources are usually centrally managed and allocated in pooling and clouding manners, including allocating resources within a system and between multiple systems of the ISAC network.
\begin{figure}[htbp]
\centerline{\includegraphics[width=0.5\textwidth,height=0.4\textwidth]{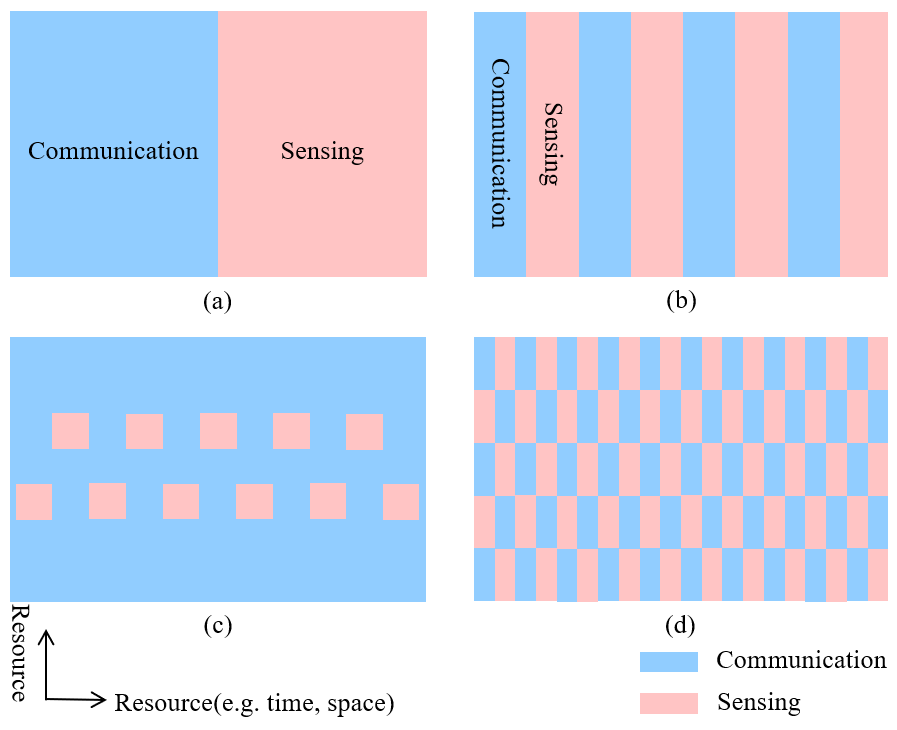}}
\vspace{-0.4em}
\caption{Types of resource allocation in ISAC.}
\label{3}
\vspace{-1em}
\end{figure}

Taking two-dimensional resource allocation as an example, it is possible to roughly summarize the four integration-level ISAC resource allocation methods mentioned in the previous section into the three forms of (a), (b), and (c) in Fig.~\ref{3}. There is an urgent need to improve the performance and efficiency of ISAC systems by appropriately allocating and scheduling resources to different tasks or applications based on demand, and to obtain higher integration gains.
Designing and selecting multi-objective optimization criteria similar to Pareto theory, game theory, etc., can facilitate the ISAC system to make effective trade-offs and decisions among multiple optimization objectives in order to obtain the best comprehensive performance, similar to that in Fig.~\ref{3} (d). Depending on the characteristics of the application scenarios, different trade-off criteria can be used for resource allocation.

\subsection{FD-ISAC: A Case for the Resource Allocation Framework}
To ensure that the communication and sensing branches work in parallel, the ISAC systems need to have the capability of simultaneous transmit and receive (STAR) for full duplex. 
Full-duplex operation has emerged as a vital factor for both SIC and effective management of sensing and communication beams in ISAC systems with resource allocation of different layers\cite{7122931}, \cite{9363029}.
The in-depth research on FD-ISAC will lead to fruitful and promising breakthroughs in the field of ISAC, so it is necessary to investigate FD-ISAC resource allocation schemes.

To achieve optimal resource allocation and mutual support in FD-ISAC systems, we consider our aforementioned multi-granularity resource allocation framework to be a promising solution.
From the hardware layer, FD can achieve SIC from propagation, analog, and digital domains for seamlessly integrating sensing and communication, while ISAC needs to realize the waveform design from time, frequency and space dimensions in the digital domain to enable the ability of STAR \cite{10035959}, \cite{s22010109},\cite{220311}, \cite{220110}. 
However, this contradiction is not invincible.
What we need to know is that the optimal allocation is not the same as achieving the maximum isolation, because high isolation can lead to increased system complexity, additional hardware overhead and increased power consumption.
For example, to achieve 90dB isolation, we can achieve 30dB isolation in each of the analog, digital and propagation domains.
However, such allocation is not fixed and will be adjusted as the dimensional condition of bandwidth changes, since it is more challenging to achieve the same isolation within larger bandwidth. 
Besides, for the mixed modes of monostatic ISAC and bistatic ISAC, the trade-off for isolation, system performance, complexity and cost must be achieved by comprehensive resource allocation in the whole ISAC network.
In conclusion, it is essential to allocate the resources in hardware and signal layers with different domains and dimensions. And also, the scale of resource allocation must be distinct for various scenarios.
The above target is exactly what our proposed multi-granularity resource allocation framework can achieve.

\section{Future Research Directions}
Although resource allocation in ISAC has been explored in multiple directions over the past years, there are still many open challenges that have not been widely investigated. In the following, we will discuss some of these open issues and future research directions in-depth.

\paragraph{AI-aided ISAC Resource Allocation}
It is difficult to solve the complex multi-system resource allocation problem in ISAC by traditional mathematical models and signal processing techniques alone, especially in networked ISAC.
The powerful data-driven AI algorithms are considered very promising in optimizing the performance and efficiency of ISAC systems. It can not only learn from historical data to continuously improve resource allocation schemes, but also intelligently perform dynamic scheduling and resource allocation based on real-time demand. In complex and data-rich ISAC application scenarios, AI-aided resource allocation deserves further exploration, and the 5G upper layer application AI approach may provide some inspiration\cite {9286851}.
%Facing the extremely complex and data-rich application scenarios of 6G, it will be a future development trend to use machine learning and other AI technologies to solve the resource allocation problem in ISAC. 
\paragraph{Practical ISAC Resource Allocation Criteria}
Resource allocation in ISAC is by nature a multi-objective optimization problem, where theories such as Pareto analysis\cite{9838610}, game theory, and Lagrangian optimization are often utilized. However, there is usually no optimal solution for multi-objective optimization problems. The selection and design of the criterion deserve more attention than the formulation of the optimization problem. 

\paragraph{Breakthrough on ISAC fundamental Theory Integration}
ISAC involves multi-system and multi-form data, which requires research on the underlying theory to achieve heterogeneous fusion. It is necessary to bridge the boundary between information theory and estimation theory and use mathematical methods to resolve the contradiction between the separated metric systems. Breaking the fundamental theoretical boundaries can help enhance the ISAC integration gain and further improve the effectiveness of resource allocation.

\section{Conclusion}
In this paper, we provide a comprehensive overview of resource allocation issues in ISAC. 
Based on the integrated properties of communication and sensing, we propose a new framework based on multi-granularity resource allocation. 
Under this framework, we discuss the multi-layer, multi-domain, multi-dimension and multi-scale resource allocation strategies.
Especially, we also give an FD-ISAC case to demonstrate the implementability of the framework.
%First, we introduce the evolutionary path and technical concerns of ISAC, followed by a discussion focusing on resource allocation techniques. 
%After that, according to the different integrated levels of the system, the resource allocation in ISAC is discussed in details, namely, resource separation, orthogonal, convergence, and collaboration phases.
%In addition, we establish a new model of resource allocation from a multi-granularity perspective which sheds light on future developments.
%Finally, we discuss several challenges and opportunities in this field.

\bibliographystyle{ieeetr} %ieeetr国际电气电子工程师协会期刊
\bibliography{ref} % ref就是之前建立的ref.bib文件的前缀。
\end{document}